\documentclass[twocolumn,epsfig,aps,prb]{revtex4}

\begin{document}

\title{Metastable states of surface plasmon vacuum near the interface between metal and nonlinear dielectric}

\author{Igor I. Smolyaninov}
\affiliation{Department of Electrical and Computer Engineering, University of Maryland, College Park, MD 20742, USA}

\date{\today}

\begin{abstract}
Zero-point fluctuations of surface plasmon modes near the interface between metal and nonlinear dielectric are shown to produce a thin layer of altered dielectric constant near the interface. This effect may be sufficiently large to produce multiple metastable states of the surface plasmon vacuum.  
\end{abstract}


\maketitle

Optical bistability in various nonlinear optical systems is a topic of great current interest because of its potential implementations in optical computing \cite{1}. The use of surface electromagnetic waves such as surface plasmon-polaritons \cite{2,3} is known to reduce considerably the optical power levels that are necessary to achieve optical bistability effects \cite{4,5}. Such a power reduction is possible because of considerable enhancement of the local optical field near the metal interfaces. Very recently a possible observation of optical bistability effect at single-photon levels due to localized plasmon excitation in nonlinear optical material-filled nanoholes in gold films has been reported \cite{6}. The experimental findings were confirmed to some degree by theoretical calculations in \cite{6,7}. This development together with other photon-blockade effects \cite{8} observed in nonlinear optical cavities at single-photon levels may open new ways of designing optical quantum computers. It justifies continued interest in nonlinear optical effects at extremely low light levels.  

As we continue to reduce number of real quanta of energy in a surface plasmon system, and still observe nonlinear optical effects, we may ask ourselves if the properties of the nonlinear dielectric near the metal interface are somehow altered due to zero-point fluctuations in the vacuum state of the surface plasmon system. In other words, can we observe nonlinear optical behavior even without a single plasmon quantum? It appears that the answer to this question should be yes. Zero-point fluctuations of the surface plasmon field $E_{vac}$ must produce some average $E^2_{vac}(z)$, which is very weak at large distances $z$ from the metal surface, but must grow exponentially towards the surface (similar to the behavior of the $E$-field in the surface plasmon wave). Let us assume that the dielectric near the metal surface exhibits third-order nonlinearity:

\begin{equation}
\label{eq1}
\epsilon _d= \epsilon ^{(1)} + 4\pi \chi ^{(3)}E^2 ,
\end{equation}

where $\epsilon ^{(1)}$ and $\chi ^{(3)}$ are the linear dielectric constant and the third order nonlinear susceptibility, respectively (here we consider the case of a central symmetric nonlinear material with $\chi ^{(2)}=0$). Far from the metal surface $\epsilon = \epsilon ^{(1)}$. However, near the metal we should see an altered layer of the material with the dielectric constant shifted by $4\pi \chi ^{(3)}E^2_{vac}$. But how large and noticeable is this effect? 

Our results indicate that this effect may be surprisingly large even in materials with modest $\chi ^{(3)}$ nonlinearities. It may be sufficiently large to produce multiple metastable states of the surface plasmon vacuum, so one can talk about optical bistability of the vacuum itself, regardless of the number of real photons or plasmons in the system. Even though surprising, this effect is not unique or extremely unusual. Metastable vacuums do occur in many field theories such as quantum chromodynamics, quantum gravity, etc. The origin of our Universe is believed to be due to a spontaneous decay of some metastable quantum vacuum state. Thus, in addition to optical computing applications, experimental study of metastable surface plasmon vacuums may provide us with a useful toy model which could facilitate our understanding of somewhat similar field-theoretical models of other quantum vacuums.    

Let us start by considering the zero-point energy of the surface plasmon vacuum of a square $a\times a$ region of a thin metal film with thickness $d<<a$ (Fig.1(a)). Let us assume that the half-spaces above and below the film are filled with the same third-order nonlinear dielectric and consider the dispersion law of a surface plasmon (SP), which propagates along the metal-dielectric interface (the SP field decays exponentially both inside the metal and the dielectric). In such a case the dispersion law can be written as \cite{9} 

\begin{equation}  
\label{eq2} 
k^2=\frac{\omega ^2}{c^2}\frac{\epsilon _d\epsilon _m(\omega )}{\epsilon _d+\epsilon _m(\omega)\pm 2\epsilon _de^{-kd}} ,
\end{equation}

where $\epsilon _m(\omega )$ is the frequency-dependent dielectric constant of the metal. If we assume that according to the Drude model $\epsilon _m=1-\omega _p^2/\omega ^2$ ($\omega _p$ is the plasma frequency of the lossless metal), and $d$ is large, the dispersion law can be simplified as

\begin{equation}  
\label{eq3} 
k^2=\frac{\omega ^2}{c^2}\frac{\epsilon _d\epsilon _m(\omega )}{\epsilon _d+\epsilon _m(\omega)} ,
\end{equation}

where $\epsilon _m(\omega )$ is real.
 This dispersion law is shown in Fig.1(b) for the cases of metal-vacuum and metal-dielectric interfaces. It starts as a "light line" in the respective dielectric at low frequencies, and approaches asymptotically $\omega _{sp}=\omega _p/(1+\epsilon _d)^{1/2}$ at large wave vectors. The latter frequency corresponds to the so-called surface plasmon resonance. The dispersion law (2) also looks very similar to Fig.1(b) in the general case of a lossy metal film if the metal film thickness $d$ is small, so that the imaginary part of the term $\pm 2\epsilon _de^{-kd}$ compensates the imaginary part of $\epsilon _m$. In such cases the plasmon wavevector $k$ also diverges \cite{9} at some frequency, which is close to $\omega _{sp}=\omega _p/(1+\epsilon _d)^{1/2}$. Since at every wavevector the SP dispersion law is located to the right of the "light line", the SPs of the plane metal-dielectric interface are decoupled from the free-space photons due to the momentum conservation law. Nevertheless, very-short wavelength (in the 10-50 nm range) plasmons with the frequencies near the frequency of surface plasmon resonance were observed in a number of near-field optical experiments \cite{10,11}.

The main contribution to the zero-point energy of the surface plasmon field comes from the frequency region near the surface plasmon resonance $\omega _{sp}$ in which $\omega $ does not depend on $k$, since the plasmon density of states is very large in this region. As a result, the total zero-point energy of the plasmon vacuum can be approximated as 

\begin{equation}
\label{eq4} 
U_{vac} \approx \frac{1}{2}\Sigma _k \hbar \omega _{sp}, 
\end{equation}

where summation has to be done over all the plasmon modes $k$ of the square region of the thin metal film under consideration. The surface plasmon eigenmodes of this square region are defined by the two-component wave vector $(k_x,k_y)=\pi/a\times (n_x,n_y)$, where $n_x$ and $n_y$ are integer. Thus, if $a$ of the order of 100-200 nm is selected, the quasistatic approximation used in equation (4) is well-justified. The summation over all possible surface plasmon wave vectors has to be cut off at some $k_{max}$, which according to \cite{12} is defined by the Landau damping and equals to the electron Fermi momentum $k_F$ so that $\mid k_{max}\mid \sim k_F=(6\pi ^2n_e)^{1/3}$ (where $n_e$ is the electron density). However, if we are less optimistic about the total zero-point energy, we may assume that $k_{max}=2\pi /\lambda _{min}$ corresponds to the shortest surface plasmon wavelength $\lambda _{min}$ observed in the experiment. 

Thus, the total number of plasmon modes on both the top and the bottom interfaces of the metal film is approximately $4k^2_{max}a^2/\pi ^2$, and 

\begin{equation}
\label{eq5} 
U _{vac} = \frac{2}{\pi ^2}\hbar \omega _{sp}k^2_{max}a^2
\end{equation}

On the other hand, zero-point energy of the plasmon field in this region can be expressed as

\begin{equation}
\label{eq6} 
U _{vac} = \frac{<E^2_{sp}>}{4\pi }a^2<1/k_{sp}> ,
\end{equation}

where $<E^2_{sp}>$ is the average of the square of the electric field of the plasmon zero-point fluctuations, and $<1/k_{sp}>=2/k_{max}$ is the average decay length of the plasmon field away from the metal surface, which is obtained by averaging $1/k$ over all the surface plasmon modes of the square film region. 
As a result, the electric field of plasmon zero-point fluctuations can be found as

\begin{equation}
\label{eq7} 
<E^2_{sp}> = \frac{4}{\pi }\hbar \omega _{sp}k^3_{max} 
\end{equation}

The shift of the dielectric constant of the nonlinear optical material near the metal surface equals to  

\begin{equation}
\label{eq8} 
\Delta \epsilon _d=4\pi \chi ^{(3)}<E^2_{sp}>=16\chi ^{(3)}\hbar \omega _{sp}k^3_{max}=A\omega _{sp} 
\end{equation}

Assuming rather cautious $\lambda _{min}=30 nm$, we obtain $\Delta \epsilon _d\sim 10^9\chi ^{(3)}$ esu$^{-1}$. Taking into account that such nonlinear dielectrics as 3BCMU and 4BCMU polydiacetylene materials have $\chi ^{(3)}$ as large as $10^{-10}$ esu \cite{13}, very large zero-field induced shifts $\Delta \epsilon _d\sim 0.1$ may be expected near the interface between metal and such nonlinear dielectrics (note that these materials were used in the earlier photon-blockade experiments \cite{6}). If less cautious upper limit $\mid k_{max}\mid \sim k_F$ is assumed, the resulting shift would be much larger ($\Delta \epsilon _d\sim 10^{14}\chi ^{(3)}$ esu$^{-1}$), which would make this effect noticeable even in regular nonlinear optical materials.    

The predicted nonlinear dielectric constant shifts are so large that the effects of bistability may be expected for the surface plasmon vacuum state. The origin of this effect would be roughly the same as the origin of bistability effects in regular surface plasmon propagation \cite{4}. Using the constant $A$ from equation (8) which is proportional to $\chi ^{(3)}$, the frequency of surface plasmon resonance may be written as 

\begin{equation}
\label{eq9} 
\omega _{sp}=\frac{\omega _p}{(1+\epsilon _d)^{1/2}}=\frac{\omega _p}{(1+\epsilon ^{(1)}+A\omega _{sp})^{1/2}}
\end{equation}

Let us initially assume that $\chi ^{(3)}$ and $A$ are constants independent of the optical frequency (this is not the case for 4BCMU polydiacetylene in which $\chi ^{(3)}$ rapidly oscillates in the visible range from 450 nm to 700 nm \cite{13}). Under such an assumption the frequency of surface plasmon resonance is given by the following cubic equation: 

\begin{equation}
\label{eq10} 
A\omega ^3_{sp}+(\epsilon ^{(1)}+1)\omega ^2_{sp}-\omega ^2_p=0
\end{equation}

The graphic solution of this cubic equation is presented in Fig.2 for the cases of positive and negative $\chi ^{(3)}$. Positive $\chi ^{(3)}$ increases the dielectric constant in the immediate vicinity of the metal surface and shifts the frequency of the surface plasmon resonance downward because of the zero-point fluctuations. Negative $\chi ^{(3)}$ has an opposite effect on the surface plasmon resonance: the dielectric constant near the metal surface decreases, which leads to increase of the frequency of the surface plasmon resonance. In addition, another meaningful root of equation (10) may appear at even higher frequencies, if this root is below the plasma frequency of metal $\omega _p$. 
Analytical solution of the cubic equation (10) indicates that this interesting behavior occurs near $A^2=4(\epsilon ^{(1)}+1)^3/27\omega ^2_p$ and the two roots are located around 

\begin{equation}
\label{eq11} 
\omega _{2,3}=\frac{3^{1/2}\omega _p}{(\epsilon ^{(1)}+1)^{1/2}}
\end{equation}

Thus, two states are possible for the surface plasmon vacuum when $\chi ^{(3)}=const<0$. The upper vacuum state is metastable and once created, it should eventually spontaneously decay into the lower-energy vacuum state. 
Since $\chi ^{(3)}$ of such materials as 4BCMU polidyacetylene is not actually constant as a function of frequency, but rather quickly oscillates in magnitude and sign in the visible frequency range \cite{13}, other scenarios of bistable or even multi-stable plasmon vacuums are possible. If $\chi ^{(3)}(\omega )$ changes sign from positive to negative in the vicinity of $\omega ^{(1)}_{sp}=\omega _p/(1+\epsilon ^{(1)})^{1/2}$, the lower-energy vacuum state from the pair obtained for $\chi ^{(3)}<0$, which is situated above $\omega ^{(1)}_{sp}$ becomes metastable, and would eventually decay into a stable state located below $\omega ^{(1)}_{sp}$ at $\chi ^{(3)}>0$. 

The metastable states of surface plasmon vacuum may be created using the effect of bistability observed experimentally in the excitation of surface plasmons over an interface between metal and nonlinear dielectric \cite{4}. Real physical surface plasmons excited in the upper energy state would create a necessary seed layer with an altered dielectric constant near the metal interface. After all the real plasmons decay or radiate (due to the losses in metal and surface roughness), the surface plasmon vacuum may stay in the metastable state for a while. Exactly how long this metastable vacuum state is going to last would be best addressed by a pump-probe type of optical experiment. Theoretical answer to this question would meet problems, which are somewhat similar to the problems in quantum gravity, neither of which has been solved so far. In order to understand this difficulty let us recall the well-known analogy between media electrodynamics and gravitation theory.   

The realization that solid-state toy models may help in an understanding of electromagnetic phenomena in curved space-time has led to considerable recent effort in developing toy models of electromagnetic \cite{14,15} and sonic \cite{16,17,18,19} black holes. In the case of media electrodynamics this may be possible because of an analogy between the propagation of light in matter and in curved space-times: it is well known that Maxwell equations in a general curved space-time background $g_{ij}(x,t)$ are equivalent to the phenomenological Maxwell equations in the presence of a matter background with nontrivial dielectric and magnetic permeability tensors $\epsilon _{ij}(x,t)$ and $\mu _{ij}(x,t)$ \cite{14}:

\begin{equation}
\label{eq12} 
\epsilon _{ij}(x,t)=\mu _{ij}(x,t)=(-g)^{1/2}\frac{ g^{ij}(x,t)}{g_{00}(x,t)}
\end{equation}

In this analogy, some features of the black hole event horizon may be emulated by a surface of singular dielectric constant and magnetic permeability, so that the speed of light goes to zero, and light is "frozen" near such a surface. 

If we accept this language, the nonlinear optical corrections to $\epsilon _{ij}(x,t)$ and $\mu _{ij}(x,t)$ of the form described by eq.(1) would describe an effective gravitational interaction between the quanta of electromagnetic field. This fact is very easy to illustrate in situations in which $\epsilon _{ij}(x,t)$ and $\mu _{ij}(x,t)$ depend only on one spatial coordinate $z$ (such as the Rindler geometry). After straightforward calculations one obtains the following expression for the $R_{00}$ component of the Riemann tensor (recall that the trace of the energy-momentum tensor $T=0$ for the electromagnetic field): 

\begin{equation}
\label{eq13} 
R_{00}=\frac{1}{\epsilon }\frac{\partial ^2}{\partial z^2}(\frac{1}{\epsilon})=\frac{16\pi \chi ^{(3)}k^2}{\epsilon ^{(1)3}}E^2=\frac{8\pi k_{eff}}{c^4}T_{00} ,
\end{equation}

which reproduces the Einstein equation with the effective gravitational constant $k_{eff}$, which is proportional to $\chi ^{(3)}$. Unfortunately, in normal three-dimensional optics situations $\epsilon ^{(1)}$ and $\chi ^{(3)}$ of optical materials are small, and no non-trivial effects of effective gravity may be emulated. All the electromagnetic black hole models consider motion of light in some static geometry of $\epsilon _{ij}(x)$ and $\mu _{ij}(x)$. 

Very recently we have pointed out that this situation changes in the case of two-dimensional optics of surface plasmon-polaritons \cite{20}. It is clear from Fig.1(b) that around the frequency of surface plasmon resonance both phase and group velocity of surface plasmons tend to zero, and the effective two-dimensional dielectric constant $\epsilon _2^{(1)}$ of the dielectric diverges as seen by the plasmons which propagate along the metal-dielectric interface. In a similar fashion the effective two-dimensional $\chi _2^{(3)}$ of the nonlinear dielectric diverges near the surface plasmon resonance: small changes of the three-dimensional dielectric constant $\epsilon _d$ due to the $4\pi \chi ^{(3)}E^2$ term in equation (1) are perceived as very large changes of the effective two-dimensional dielectric constant by surface plasmon-polaritons. Thus, effective gravity of surface plasmons becomes very strong. As has been described in \cite{20}, gravitational collapse of surface plasmon field can be emulated in a self-focusing experiment performed near the frequency of the surface plasmon resonance. However, such self-focusing of surface plasmons emulates only the effects of classic general relativity. 

In order to emulate quantum gravity one would need to make plasmons behave as the quanta of the effective gravitational field. This is easy to achieve in a non-symmetric nonlinear optical medium with non-zero $\chi ^{2}$ susceptibility.
In such media  

\begin{equation}
\label{eq14}
\epsilon _d= \epsilon ^{(1)}+4\pi \chi ^{(2)}E + 4\pi \chi ^{(3)}E^2 ,
\end{equation}

where the tensor indices of $\epsilon $, $\chi ^{2}$, $\chi ^{3}$, and $E$ are omitted for the sake of simplicity. Comparison of equations (12) and (14) indicates that oscillations of the effective gravitational metric are proportional to the oscillations of the $E$ field of the surface plasmon waves. In effect, surface plasmons become effective gravitons of the system. Since the inversion symmetry is broken near every interface, surface $\chi ^{2}$ terms are bound to appear at the interface of metal with any nonlinear optical medium, so that the equation (14) correctly describes nonlinear behavior of any such interface.

To complete our quantum gravity analogy let us derive the effective Planck scale $L_{pl}^{eff}=(\hbar k_{eff}/c^3)^{1/2}$ of our system. Taking into account the value of the effective gravitational constant $k_{eff}$ from equation (13), we find   

\begin{equation}
\label{eq15}
L_{pl}^{eff}=(\frac{\hbar k_{eff}}{c^3})^{1/2}=(\frac{8\pi \hbar c\chi ^{(3)}k^2}{\epsilon ^{(1)3}})^{1/2}
\end{equation}

Assuming $\chi ^{(3)}\sim 10^{-10}$ esu, and the plasmon wavelength $\lambda =30$ nm, which defines $k=2\pi /\lambda $, we obtain the effective Planck scale as $L_{pl}^{eff}=2$ nm, while at $\lambda =10$ nm we get $L_{pl}^{eff}\approx \lambda $. These simple arguments indicate that the effects of effective quantum gravity are very strong in our system. Since the lifetime of the metastable surface plasmon vacuum is determined by the quantum fluctuations of the effective metric $\epsilon _d$ of the nonlinear optical material, theoretical resolution of this problem seems to be difficult at the moment. On the other hand, this lifetime may be measured experimentally in a toy "big bang"-like experiment. Such experiments would be very beneficial both from the point of view of designing optical computers, and from the point of view of advancing theoretical models of quantum gravity.

This work has been supported in part by the NSF grant ECS-0304046.

\begin{figure}
\begin{center}
\end{center}
\caption{(a) Model geometry for the calculations of zero-point energy of surface plasmon vacuum in the case of a square $a\times a$ region of a thin metal film with thickness $d<<a$. The half-spaces above and below the film are filled with the third-order nonlinear dielectric. (b) Surface plasmon dispersion law for the cases of metal-vacuum and metal-dielectric interfaces. At large plasmon wavevectors $k$ the surface plasmon frequency does not depend on $k$. Nonlinear shifts of the surface plasmon resonance are shown for the cases of $\chi ^{(3)}>0$ and $\chi ^{(3)}<0$.}
\end{figure}

\begin{figure}
\begin{center}
\end{center}
\caption{ Graphic solution of equation (10) which defines the frequency of surface plasmon resonance for the cases of positive and negative $\chi ^{(3)}$ of the nonlinear optical material. The roots are shown by black dots. If $\chi ^{(3)}$ changes sign near $\omega _{sp}$ all roots become metastable states of surface plasmon vacuum.}
\end{figure}

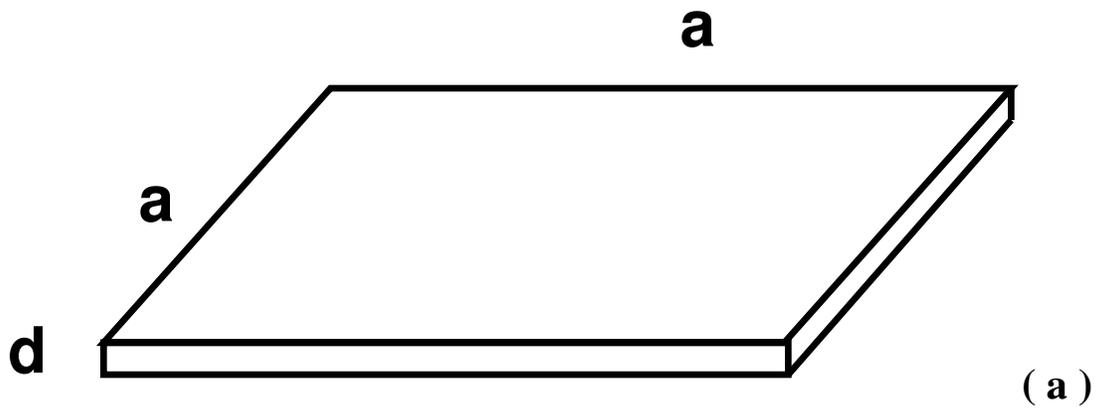
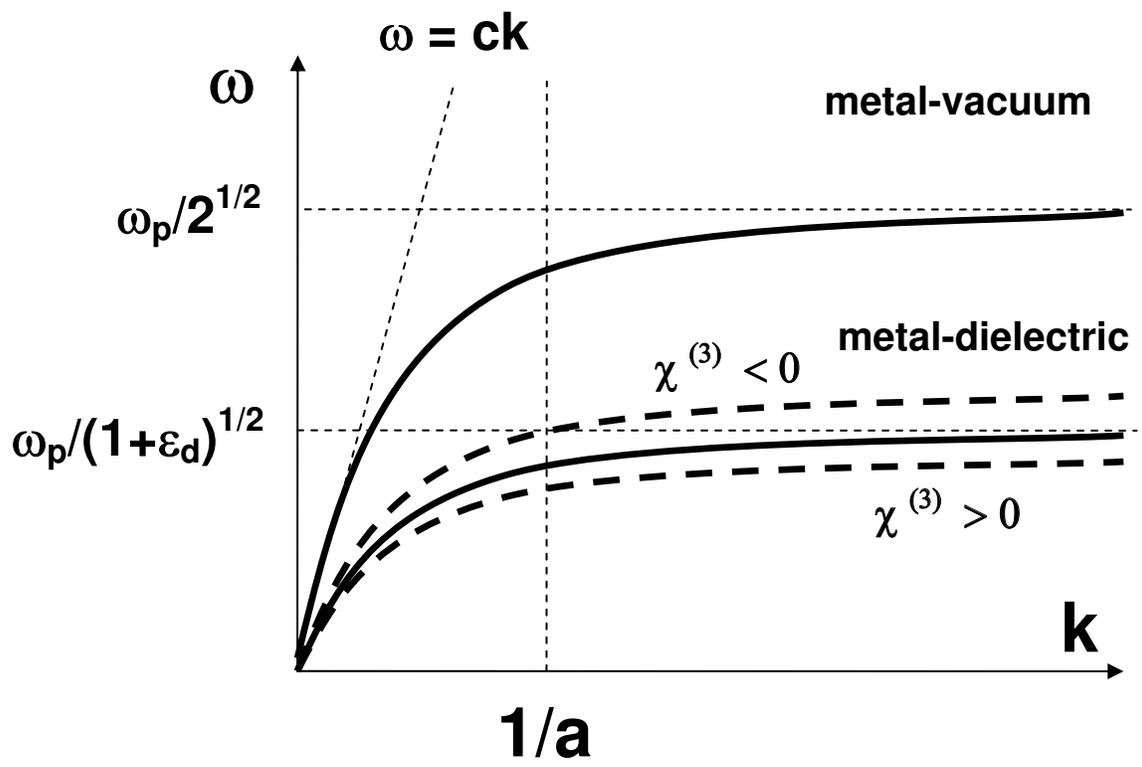
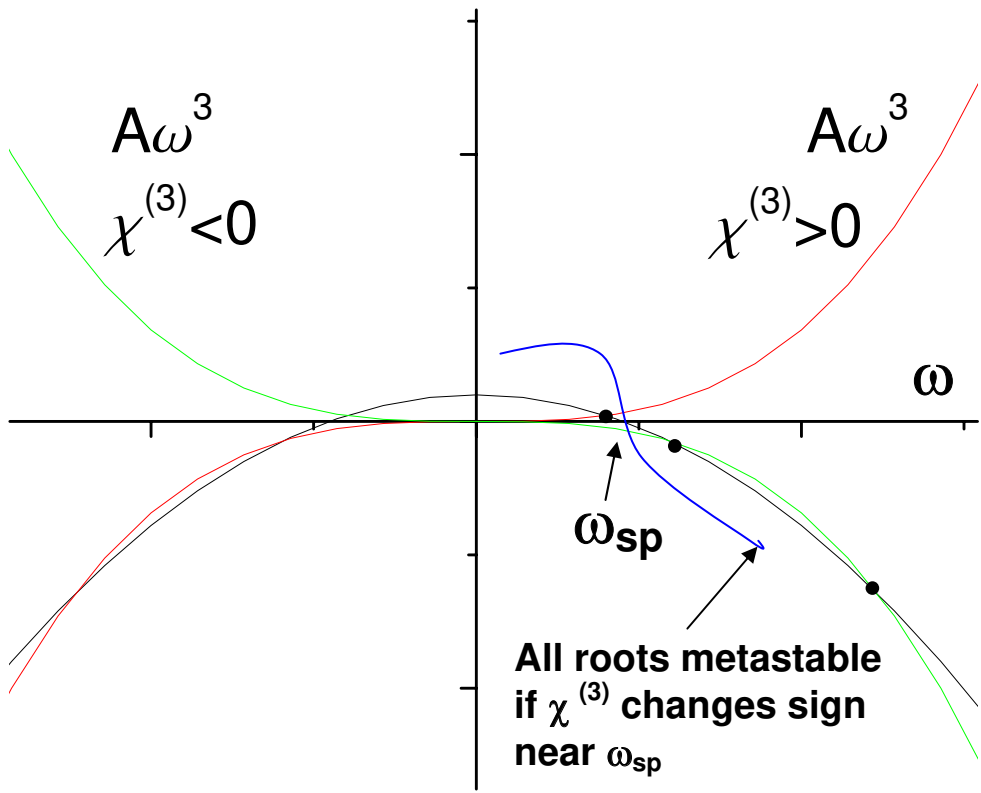
\end{document}